\documentclass[aps,prl,preprint,tightenlines,superscriptaddress,showpacs,byrevtex]{revtex4}
\def\to{\rightarrow}

\usepackage{graphicx} 
\usepackage{dcolumn}  

\graphicspath{{ps}}

\begin{document}


\preprint{\vbox{ \hbox{   }
                 \hbox{BELLE-CONF-0510}
                 \hbox{LP2005-145}
                 \hbox{EPS05-482} 
}}

\title{ \quad\\[0.5cm]    
 Evidence for the decay $\Upsilon(4S) \rightarrow \Upsilon(1S) \pi^{+} \pi^{-} $
}

\affiliation{Aomori University, Aomori}
\affiliation{Budker Institute of Nuclear Physics, Novosibirsk}
\affiliation{Chiba University, Chiba}
\affiliation{Chonnam National University, Kwangju}
\affiliation{University of Cincinnati, Cincinnati, Ohio 45221}
\affiliation{University of Frankfurt, Frankfurt}
\affiliation{Gyeongsang National University, Chinju}
\affiliation{University of Hawaii, Honolulu, Hawaii 96822}
\affiliation{High Energy Accelerator Research Organization (KEK), Tsukuba}
\affiliation{Hiroshima Institute of Technology, Hiroshima}
\affiliation{Institute of High Energy Physics, Chinese Academy of Sciences, Beijing}
\affiliation{Institute of High Energy Physics, Protvino}
\affiliation{Institute of High Energy Physics, Vienna}
\affiliation{Institute for Theoretical and Experimental Physics, Moscow}
\affiliation{J. Stefan Institute, Ljubljana}
\affiliation{Kanagawa University, Yokohama}
\affiliation{Korea University, Seoul}
\affiliation{Kyoto University, Kyoto}
\affiliation{Kyungpook National University, Taegu}
\affiliation{Swiss Federal Institute of Technology of Lausanne, EPFL, Lausanne}
\affiliation{University of Ljubljana, Ljubljana}
\affiliation{University of Maribor, Maribor}
\affiliation{University of Melbourne, Victoria}
\affiliation{Nagoya University, Nagoya}
\affiliation{Nara Women's University, Nara}
\affiliation{National Central University, Chung-li}
\affiliation{National Kaohsiung Normal University, Kaohsiung}
\affiliation{National United University, Miao Li}
\affiliation{Department of Physics, National Taiwan University, Taipei}
\affiliation{H. Niewodniczanski Institute of Nuclear Physics, Krakow}
\affiliation{Nippon Dental University, Niigata}
\affiliation{Niigata University, Niigata}
\affiliation{Nova Gorica Polytechnic, Nova Gorica}
\affiliation{Osaka City University, Osaka}
\affiliation{Osaka University, Osaka}
\affiliation{Panjab University, Chandigarh}
\affiliation{Peking University, Beijing}
\affiliation{Princeton University, Princeton, New Jersey 08544}
\affiliation{RIKEN BNL Research Center, Upton, New York 11973}
\affiliation{Saga University, Saga}
\affiliation{University of Science and Technology of China, Hefei}
\affiliation{Seoul National University, Seoul}
\affiliation{Shinshu University, Nagano}
\affiliation{Sungkyunkwan University, Suwon}
\affiliation{University of Sydney, Sydney NSW}
\affiliation{Tata Institute of Fundamental Research, Bombay}
\affiliation{Toho University, Funabashi}
\affiliation{Tohoku Gakuin University, Tagajo}
\affiliation{Tohoku University, Sendai}
\affiliation{Department of Physics, University of Tokyo, Tokyo}
\affiliation{Tokyo Institute of Technology, Tokyo}
\affiliation{Tokyo Metropolitan University, Tokyo}
\affiliation{Tokyo University of Agriculture and Technology, Tokyo}
\affiliation{Toyama National College of Maritime Technology, Toyama}
\affiliation{University of Tsukuba, Tsukuba}
\affiliation{Utkal University, Bhubaneswer}
\affiliation{Virginia Polytechnic Institute and State University, Blacksburg, Virginia 24061}
\affiliation{Yonsei University, Seoul}
  \author{K.~Abe}\affiliation{High Energy Accelerator Research Organization (KEK), Tsukuba} 
  \author{K.~Abe}\affiliation{Tohoku Gakuin University, Tagajo} 
  \author{I.~Adachi}\affiliation{High Energy Accelerator Research Organization (KEK), Tsukuba} 
  \author{H.~Aihara}\affiliation{Department of Physics, University of Tokyo, Tokyo} 
  \author{K.~Aoki}\affiliation{Nagoya University, Nagoya} 
  \author{K.~Arinstein}\affiliation{Budker Institute of Nuclear Physics, Novosibirsk} 
  \author{Y.~Asano}\affiliation{University of Tsukuba, Tsukuba} 
  \author{T.~Aso}\affiliation{Toyama National College of Maritime Technology, Toyama} 
  \author{V.~Aulchenko}\affiliation{Budker Institute of Nuclear Physics, Novosibirsk} 
  \author{T.~Aushev}\affiliation{Institute for Theoretical and Experimental Physics, Moscow} 
% ...
\collaboration{Belle Collaboration}
\noaffiliation

\begin{abstract}
A study of transitions between $\Upsilon$ states with the emission of
charged pions using 398 fb$^{-1}$ of data collected with the Belle 
detector at the KEKB asymmetric energy $e^+e^-$ collider is presented.
A clear peak from the decay $\Upsilon(1S)\to\mu^+\mu^-$
is observed in the invariant mass distribution of 
$(\mu^+ \mu^-)$ pairs from the $(\mu^+ \mu^- \pi^+ \pi^- X)$  event sample.
The mass difference distribution
($M_{\mu^+\mu^-\pi^+\pi^-}-M_{\mu^+\mu^-}$) for $M_{\mu^+\mu^-}$ from
the $\Upsilon(1S)$ mass region has two peaks from
$\Upsilon(2S,3S)\to \Upsilon(1S)\pi^+\pi^-$ decays, with no
background.
A third peak at $\Delta M =  (1119.0\pm 1.4)$~MeV/$c^2$ 
can be interpreted as evidence of a signal from the decay
$\Upsilon(4S)\to \Upsilon(1S)\pi^+\pi^-$
with a subsequent $ \Upsilon(1S) \rightarrow \mu^+ \mu^-$ transition. 
This is the first example of a non-$B \bar{B}$ decay 
of the $\Upsilon(4S)$ resonance.
The preliminary estimated branching fraction is equal to
$\mathcal{B}(\Upsilon(4S)\to \Upsilon(1S)\pi^+\pi^-) = 
(1.1 \pm 0.2(\mathrm{stat.}) \pm 0.4(\mathrm{sys.}))\times 10^{-4}.$
\end{abstract}


\maketitle

\tighten

{\renewcommand{\thefootnote}{\fnsymbol{footnote}}}
\setcounter{footnote}{0}
\section{Introduction}
The bottomonium state $\Upsilon(4S)$ has a mass above the
threshold for  $B \bar B$ pair
production and decays mainly into these $B$-meson pairs
($\mathcal{B} ( \Upsilon(4S) \rightarrow B \bar B)\!>$96\%~\cite{PDG}).
The decay modes
$\Upsilon(4S) \rightarrow \Upsilon(mS) \pi \pi$ with $m=$1, 2, 3 
should exist. These decays are analogous to the decay modes
of the low lying $\Upsilon$ states~\cite{theory}.
Upper limits on the branching fractions of 
$\Upsilon(4S) \rightarrow \Upsilon(1S) \pi^+ \pi^-$ and 
$\Upsilon(4S) \rightarrow \Upsilon(2S) \pi^+ \pi^-$  decays have been 
set by the CLEO experiment~\cite{CLEO_U4S}.
Similar decay modes of the charmonium state $\psi(3770)$,
which has a mass above the $D \bar{D}$ production threshold 
have been observed recently~\cite{charm}.

In this paper we report the first evidence for the decay mode
 $\Upsilon(4S) \rightarrow \Upsilon(1S) \pi^+ \pi^-$ 
from the Belle experiment.   

\section{Event selection}
In this study 398  fb$^{-1}$ of data collected by the Belle detector~\cite{Belle}
on the $\Upsilon(4S)$ resonance and in the nearby continuum is used. 
Well reconstructed charged
particles and photons are used to reconstruct the decay
$\Upsilon(4S) \rightarrow \Upsilon(1S) \pi^+ \pi^-$ with the subsequent
leptonic decay $\Upsilon(1S)\to \ell^+ \ell^-$.

Several selection criteria for charged tracks and
neutral particles in an event were applied.

Charged particle candidates were required to satisfy the
following requirements:
\begin{itemize}
\item transverse momentum, $p_{\mathrm{t}} \!>$30 MeV/$c$;
\item impact parameter transverse to the beam, $dr\!<$3 cm;
\item impact parameter along the beam axis,  $dz\!<$4 cm;
\item tracks that are not identified as a decay product
of a reconstructed secondary V$^0$.
\end{itemize}

Charged particles are identified by combining responses from 
the CDC, TOF and ACC subdetectors of Belle~\cite{Belle}
into a likelihood  $\cal L$$_i$~\cite{EID,MUID}
where $i$ indicates the particle type  ($e$, $\mu$, $\pi$, $K$, $p$).
A charged particle  is identified as an electron if
the corresponding likelihood ratio, $P_e\!>$0.9~\cite{EID}, 
or if the electron mass hypothesis has the highest probability.
The electron detection efficiency for  $P_e\!>$0.9
is approximately 90\% for single electrons embedded onto hadronic
events and uniformly distributed over the polar angle range 
$35^{\circ} \le \theta \le 125^{\circ}$  and the momentum range
0.5 GeV/$c \le p \le$3.0 GeV/$c$.
Charged particles are identified as muons if the corresponding 
muon likelihood ratio  is $P_{\mu} \!>$0.8.
The muon detection efficiency above a given likelihood threshold
is approximately 91.5\% for the single-track simulated muons
uniformly distributed over the polar angle range 
$20^{\circ} \le \theta \le 155^{\circ}$  and the momentum range
0.7 GeV/$c \le p \le$3.0 GeV/$c$.
Charged particles that are not identified as muons
or have a likelihood ratio $P_e\!<$0.05 were considered as pions.  

The identification of $\gamma$'s and $\pi^0$'s is based on information
from the electromagnetic calorimeter. 
Calorimeter clusters not associated with reconstructed charged tracks
are considered as $\gamma$ candidates. In addition, these
electromagnetic clusters must have energies greater than 50~MeV
and not satisfy the  $\pi^0$ hypothesis when combined with another
photon in the event.

Two electromagnetic showers with an invariant mass
 $|M_{\gamma\gamma}-m_{\pi^0}|\!<$8 MeV/$c^2$ 
(where $m_{\pi^0}$ is the nominal $\pi^0$ mass) 
form a $\pi^0 \ $  
if the confidence level for their kinematic fit to the $\pi^0 \ $ hypothesis 
is greater than 0.1 and if their reconstructed momentum
is greater than 100~MeV/$c$.

Hadronic events are selected by the standard Belle hadronic
selection. The most important of these selection criteria are the following:
multiplicity of charged tracks in an event $N_{\rm ch}\!\ge$3;
the electromagnetic calorimeter cluster multiplicity $nECL\!>$1;  
the event's visible energy $E_{\rm vis}\! \ge\!0.2\sqrt{s}$, where $\sqrt{s}$
is a center of mass energy;
the energy sum of good cluster energies in the electromagnetic
calorimeter must satisfy   0.18$\le\!E_{\rm sum}/\sqrt{s}\!\le$0.8;
the sum of the $z$ components of each charged track's and photon's momenta 
is required to satisfy $|P_{\rm z}|\!<\!0.5\sqrt{s}$. 

Events with a pair of oppositely charged particles and with
an invariant mass $M_{\mathrm{ch}^+\mathrm{ch}^-}\!>$9~GeV/$c^2$
were selected from the
hadronic sample. The invariant mass distribution for these events
in the on-resonance sample is shown in Fig. 1. The $M_{\mu^+\mu^-}$ 
invariant mass distribution when both particles are muons is also
shown in Fig. 1.  An enhancement near the $\Upsilon(1S)$ mass can
be observed.

The peak at the invariant mass $\sim$10.6 GeV/$c^2$ originates from
the process $e^+e^- \rightarrow \mu^+\mu^-(e^+e^- (\gamma))$.
About 60\% of the events in the  $M_{\mathrm{ch}^+ \mathrm{ch}^-}$
spectrum come from identified $\mu^+\mu^-$ pairs. 
The absence of $e^+ e^-$ events with
high invariant mass is mainly due to  selection criterion
for the energy sum of good cluster energies in the electromagnetic
calorimeter $E_{\rm sum}/\sqrt{s}\!\le$0.8.
Therefore we restrict our attention to $\mu^+ \mu^-$ pairs.

To reduce the background from poorly reconstructed events
we impose an additional requirement on the visible energy of selected events: 
10.5~GeV~$\!<\!E_{\mathrm{vis}}\!<$12.5~GeV.
After all cuts described above about 112.5k events were selected.
  
\section{Analysis}
To reconstruct $\Upsilon(4S)\to\Upsilon(1S) \pi^+\pi^-$ decays,
which in addition to the decay products of the $\Upsilon(1S)$ contain
two charged pions in the final state, $\mu^+ \mu^- \pi^+ \pi^- X$
events were considered.
This additional selection criterion reduces the primary sample
of events considerably (by about a factor of 100).
The number of selected  events is 957.
The muon pair invariant mass distribution, $M_{\mu^+ \mu^-}$,
for the $\mu^+ \mu^- \pi^+ \pi^- X$ events is shown in Figs.~1 and 2.
The clear low background signal for $\Upsilon(1S) \to \mu^+ \mu^-$
is shown in Fig. 2.

The fit to this distribution using a sum of the Crystal Ball 
function~\cite{CBf} for the signal
and a polynomial function for the background results in the peak position
$M_{\mu^+\mu^-}$(peak)=(9454.7$\pm$ 5.5) MeV/$c^2$, which is consistent
with the nominal $\Upsilon(1S)$ mass value
$M_{\Upsilon(1S)}$=(9460.30$\pm$ 0.26) MeV/$c^2$~\cite{PDG}.

To observe resonance states that decay into the  
$\Upsilon(1S) \ \pi^+ \pi^-$ final state
the distribution of the mass difference 
$\Delta M = (M_{\mu^+\mu^-\pi^+\pi^-} - M_{\mu^+\mu^-})$ 
where $M_{\mu^+\mu^-}$ is restricted to 
$|M_{\mu^+ \mu^-}-M_{\Upsilon(1S)}|\!<$6 MeV/$c^2$ 
was examined (see Fig. 3).
Here $M_{\Upsilon(1S)}$ is the nominal $\Upsilon(1S)$ mass.
To reduce background from $\Upsilon(1S)$ production in radiative 
return processes~\cite{rad} with the subsequent conversion of the emitted 
photon as an electron-positron pair
($\gamma* \to e^+ e^-$),
which is then misidentified as $\pi^+ \pi^-$, we impose an
additional requirement on the angle between the pion momenta
in the lab system,  cos$\theta_{\pi\pi}\!<$0.95, on the events shown in
Fig. 3.
The  cos$\theta_{\pi\pi}$ distributions for the events in the
peaks and in the sideband regions are shown in Fig. 4.
The peaks near cos$\theta_{\pi\pi}=1$ are from the radiative
return $\Upsilon(1S)$ background described above. The additional
cos$\theta_{\pi\pi}$ requirement removes this background.

Three peaks are seen in the $\Delta M$ distribution (Fig. 3). 
The first (second, third) peaks have values of 
$\Delta M \sim 0.56 (0.89, 1.12)$~GeV/$c^2$, respectively.

The first and second peaks have little or no background.
They originate from the decays 
$\Upsilon(2S) \rightarrow \Upsilon(1S) \pi^+ \pi^-$ and 
$\Upsilon(3S) \rightarrow \Upsilon(1S) \pi^+ \pi^-$
with a subsequent $ \Upsilon(1S) \rightarrow \mu^+ \mu^-$ transition,
respectively.
Fits to the first two peaks using a Gaussian (see Fig. 5a,b)
result in the following values of the peak positions
$\Delta M(1{\mathrm s}{\mathrm t})~=~(562.0~\pm~0.1)$~MeV/$c^2$,
$\Delta M(2{\mathrm n}{\mathrm d})~=~(893.5~\pm~0.2)$~MeV/$c^2$.
The mass difference between $\Upsilon(2S)$, $\Upsilon(3S)$
and the $\Upsilon(1S)$ states  presented in  PDG~\cite{PDG} is
($M_{\Upsilon(2S)}-M_{\Upsilon(1S)})=(563.0\pm 0.4)$~MeV/$c^2$,
($M_{\Upsilon(3S)}-M_{\Upsilon(1S)})=(894.9\pm 0.6)$~MeV/$c^2$, respectively.
The values obtained from the fits to the distributions in Fig. 5a,b 
by Gaussian functions are compatible with the corresponding PDG values. 
The $\chi^2$/NDF for the first and second peaks are 1.4 and 1.8,
respectively. The small differences between the mean values
obtained from the Gaussian fits and the PDG values can be explained
by systematics due to the imperfect modeling of the $\Delta M$
distributions.
We conclude that the first and second peaks are produced by the decays 
$\Upsilon(2S) \rightarrow \Upsilon(1S) \pi^+ \pi^-$ and 
$\Upsilon(3S) \rightarrow \Upsilon(1S) \pi^+ \pi^-$.

\section{Study of the third peak}
In contrast to the first two peaks, the third peak has a considerably
larger background.  
The position of the peak is derived from a fit to the distribution in Fig. 6
using a Gaussian for the signal and a polynomial function for the background to be,
 $\Delta M =  (1119.0\pm 1.4)$~MeV/$c^2$, which is in good agreement 
with the mass difference 
 $(M_{\Upsilon(4S)}-M_{\Upsilon(1S)})=(1120.0\pm 3.5)$~MeV/$c^2$
from the PDG~\cite{PDG}.
The signal above background is determined
from the fit to be, $N_{\rm ev}= (38.0\pm 6.9)$, with a statistical
significance of 7.3 standard deviations.
This peak is interpreted as a signal from the decay 
$\Upsilon(4S) \rightarrow \Upsilon(1S) \pi^+ \pi^-$
with a subsequent $ \Upsilon(1S) \rightarrow \mu^+ \mu^-$ transition. 

Using the off-resonance sample ($\sqrt{s}$=10.52 MeV, integrated luminosity
$\int \mathcal{L} dt \simeq$ 40 fb$^{-1}$), the mass difference distribution
shown in Fig. 7 shows only two peaks, which are  from
$\Upsilon(2S)$, $\Upsilon(3S)$ decays.
The enhancement at $\Delta M =$1060 MeV/$c^2$ in this distribution  
is compatible with the background level near the third peak in the on-resonance
sample (Fig. 3) and lower than the luminosity scaled yield of
the third peak in the on-resonance data.
For the renormalized on-resonance sample the total number of events 
and background in the interval 
1.11~GeV/$c^2$~$\!<$~$\!\Delta M \!<$1.135 GeV/$c^2$ corresponding to
the third peak is 
$N_{\rm tot}^{\rm res}=(5.4\pm0.8)$, $N_{\rm bkg}^{\rm res}
=(1.1\pm0.4)$
respectively. The number of events in the shifted interval
for the off-resonance sample is $N_{\rm tot}^{\rm cont}=(2.0\pm1.4)$. 
Unfortunately the low statistics of the off-resonance sample 
preclude using it to subtract background in the on-resonance
sample.

However, additional information can be obtained from the study of the 
$\pi^+ \pi^-$ system.
The invariant mass distribution of the $\pi^+\pi^-$-system
$M_{\pi^+\pi^-}$ for events from the observed peaks and background
is shown in Figs. 8a-d. 
These distributions are directly compared with model predictions
as the efficiency corrections are negligible. The background, which is small,
is not subtracted from  the $M_{\pi^+\pi^-}$ distributions 
for the  three observed peaks (Figs.~8a-c). 

The $M_{\pi^+\pi^-}$ distribution for the first peak 
($\Upsilon(2S)\to \Upsilon(1S)\pi^+\pi^-$ decay) is well described 
by the Yan model~\cite{Yan}, where the hadronic transition between heavy 
quarkonia is considered as a two-step process: the emission of gluons from
heavy quarks and subsequent conversion of these gluons into light hadrons. 
This transition can be described in the context of a ``multipole'' expansion
scheme where the gluon fields are expanded in a multipole series.

The $\Upsilon(3S)\to \Upsilon(1S)\pi^+\pi^-$ decay distribution (Fig. 8b)
on the other hand is described by the Moxhay model~\cite{Moxhay}.
In this model the decay proceeds through coupling to 
$B \bar B$, $B^* \bar B^*$, ..., intermediate states~\cite{Tuan}.

This significant coupling to the flavored sector is motivated by the fact,
that the $\Upsilon(3S)$ lies closer to the open $B\bar{B}$ production threshold
than the $\Upsilon(2S)$.
This decay mechanism causes a double peak structure in the 
$M_{\pi^+\pi^-}$ distribution.  The multipole and coupled-channel 
amplitudes interfere in the Moxhay model.

The above models were already used successfully by the CLEO
experiment~\cite{CLEO} to describe the $M_{\pi^+ \pi^-}$ distributions in the 
$\Upsilon(2S,3S)\to \Upsilon(1S)\pi^+\pi^-$ decays.
The $M_{\pi^+\pi^-}$ distribution for the decay
$\Upsilon(4S)\to \Upsilon(1S)\pi^+\pi^-$ (Fig. 8c) is better described 
by the Yan model.

The distributions for all resonance decays (Fig. 8a-c) show an enhancement at
high masses.
In contrast in the $M_{\pi^+\pi^-}$ distribution for the background (events
out of peak ranges, i.e. from the mass difference ranges 
 0.3~GeV/$c^2\!<\!\Delta M\!<$0.555~GeV/$c^2$,  
0.57~GeV/$c^2\!<\!\Delta M\!<$0.885~GeV/$c^2$, 0.905~GeV/$c^2\!<\!\Delta M\!<$1.11~GeV/$c^2$, 
1.135~GeV/$c^2\!<\!\Delta M\!<$1.4~GeV/$c^2$) 
the high mass region is significantly suppressed (see Fig. 8d).
This difference in the behaviour of the $M_{\pi^+\pi^-}$ distribution
is an additional argument in favour of a resonance interpretation
for the third peak.

The branching fraction for the 
$\Upsilon(4S)\to \Upsilon(1S)\pi^+\pi^-$ decay was extracted from
$\mathcal{B}(\Upsilon(4S) \to \Upsilon(1S)\pi^+\pi^-) = N_{\rm ev}/(N_{\rm ev}^{\rm tot}
\cdot \varepsilon \cdot \mathcal{B}(\Upsilon(1S) \rightarrow \mu^+ \mu^-)).$ 
The total number of  $\Upsilon(4S)$ in the data sample is 
$ N_{\rm ev}^{\rm tot} = (386 \pm 5) \times 10^6$,  
the nominal branching fraction 
$\mathcal{B}(\Upsilon(1S) \rightarrow \mu^+ \mu^-)=2.48$\%.
The EvtGen event generator~\cite{EvtGen} with a matrix element 
taking into account particle spins was used for the Monte Carlo simulation
of $\Upsilon(4S)\to \Upsilon(1S)\pi^+\pi^- \to \mu^+ \mu^-\pi^+\pi^-$
events. 
The Monte Carlo generated events were then passed through the detector
simulation and reconstruction programs.
The efficiency of the event reconstruction is then obtained from the Monte
Carlo sample to be $\varepsilon=0.035$.
The systematic uncertainty on the reconstruction efficiency is estimated by
comparing the number of reconstructed $\Upsilon(2S,3S)$ events
with the number of 
events that were calculated for the radiative return process
$e^+e^- \to e^+e^- \gamma \to \Upsilon(2S,3S) \gamma$ in the
model~\cite{senja}.
The difference is 35\% and is a conservative estimate of the event
reconstruction uncertainty. 
Another systematic uncertainty comes from the poor knowledge of the 
$\Upsilon(4S)\to \Upsilon(1S)\pi^+\pi^- \to \mu^+ \mu^-\pi^+\pi^-$ decay
matrix element. This was estimated from a comparison of the previously
calculated efficiency $\varepsilon$ with the efficiency calculated with a
phase space matrix element and this systematic uncertainty results in 8\%.

The preliminary result for the branching fraction is
$\mathcal{B}(\Upsilon(4S)\to \Upsilon(1S)\pi^+\pi^-) = 
(1.1 \pm 0.2(\mathrm{stat.}) \pm 0.4(\mathrm{syst.}) )\times 10^{-4}.$
The width, $\Gamma$, of the decay $\Upsilon(4S)\to \Upsilon(1S)\pi^+\pi^-$
is (2.2$\pm$1.0)~keV, which is comparable with  
$\Gamma(\Upsilon(2S)\to \Upsilon(1S)\pi^+\pi^-) =(8.1\pm 1.2)$~keV
and $\Gamma(\Upsilon(3S)\to \Upsilon(1S)\pi^+\pi^-) =(1.2\pm 0.2)$~keV.

\bigskip 
\bigskip 

\section{Conclusions}
A study of transitions between $\Upsilon$ states with the emission of
charged pions at Belle has been performed.
A clear peak from the $\Upsilon(1S)\to\mu^+\mu^-$ decay
in the invariant mass distribution of $(\mu^+ \mu^-)$ pairs from the
$\mu^+ \mu^- \pi^+ \pi^- X$  event sample is observed. 
The mass difference distribution
($M_{\mu^+\mu^-\pi^+\pi^-}-M_{\mu^+\mu^-}$) for $M_{\mu^+\mu^-}$ from
the $\Upsilon(1S)$ mass region has two peaks from 
`$\Upsilon(2S,3S)\to \Upsilon(1S)\pi^+\pi^-$ decays with no
background. 
A third peak at $\Delta M =  (1119.0\pm 1.4)$~MeV/$c^2$ is
interpreted as evidence of a signal from the decay
$\Upsilon(4S)\to \Upsilon(1S)\pi^+\pi^-$
with a subsequent $ \Upsilon(1S) \rightarrow \mu^+ \mu^-$ transition. 
The preliminary estimated branching fraction results in
$\mathcal{B}(\Upsilon(4S)\to \Upsilon(1S)\pi^+\pi^-) = 
(1.1 \pm 0.2(\mathrm{stat.}) \pm 0.4(\mathrm{sys.}))\times 10^{-4}.$
\section{Acknowledgements} 
We thank the KEKB group for the excellent operation of the
accelerator, the KEK cryogenics group for the efficient
operation of the solenoid, and the KEK computer group and
the National Institute of Informatics for valuable computing
and Super-SINET network support. We acknowledge support from
the Ministry of Education, Culture, Sports, Science, and
Technology of Japan and the Japan Society for the Promotion
of Science; the Australian Research Council and the
Australian Department of Education, Science and Training;
the National Science Foundation of China under contract
No.~10175071; the Department of Science and Technology of
India; the BK21 program of the Ministry of Education of
Korea and the CHEP SRC program of the Korea Science and
Engineering Foundation; the Polish State Committee for
Scientific Research under contract No.~2P03B 01324; the
Ministry of Science and Technology of the Russian
Federation; the Ministry of Higher Education, 
Science and Technology of the Republic of Slovenia;  
the Swiss National Science Foundation; the National Science Council and
the Ministry of Education of Taiwan; and the U.S.\
Department of Energy.

%
%

\newpage
\vspace*{2.cm}
\begin{figure}[!htb]
\includegraphics[width=0.8\textwidth]{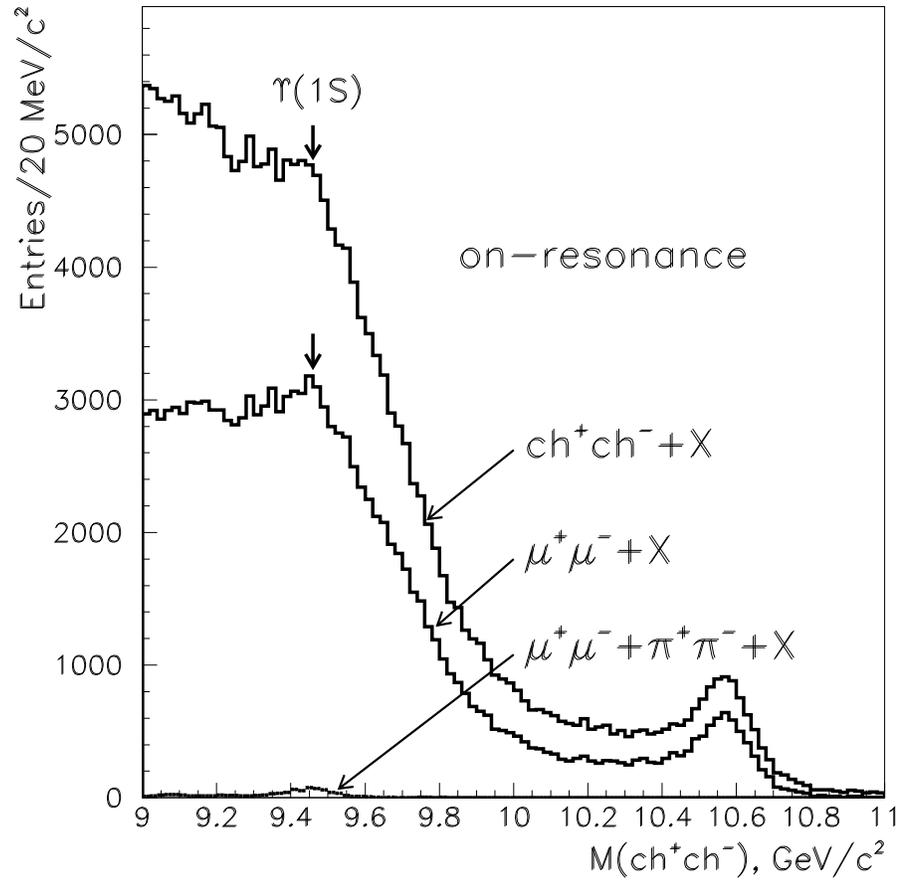}
\caption{The invariant mass distribution of charged particle pairs 
$M_{\mathrm{ch}^+\mathrm{ch}^-}$ for the standard Belle hadronic 
on-resonance sample.}
\label{fg01}
\end{figure}

\newpage
\begin{figure}[!htb]
\includegraphics[width=0.45\textwidth]{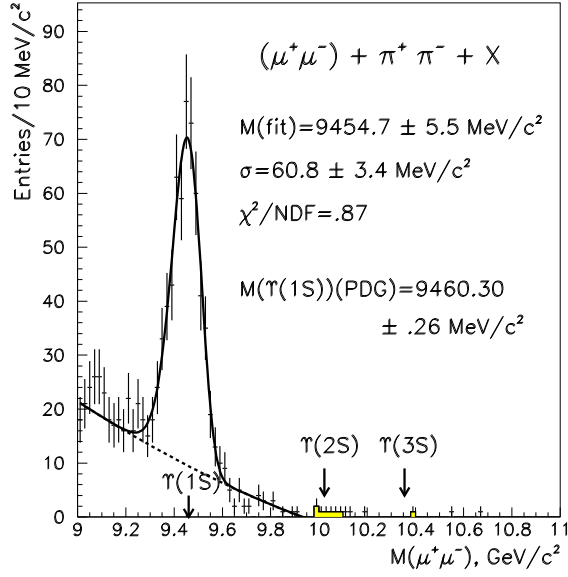}
\caption{The invariant mass distribution of muon pairs $M_{\mu^+\mu^-}$ in
$\mu^+\mu^-\pi^+ \pi^-X$ events fitted with the sum of a Crystal Ball and
polynomial function.}
\label{fg02}
\end{figure}

\begin{figure}[!htb]
\includegraphics[width=0.55\textwidth]{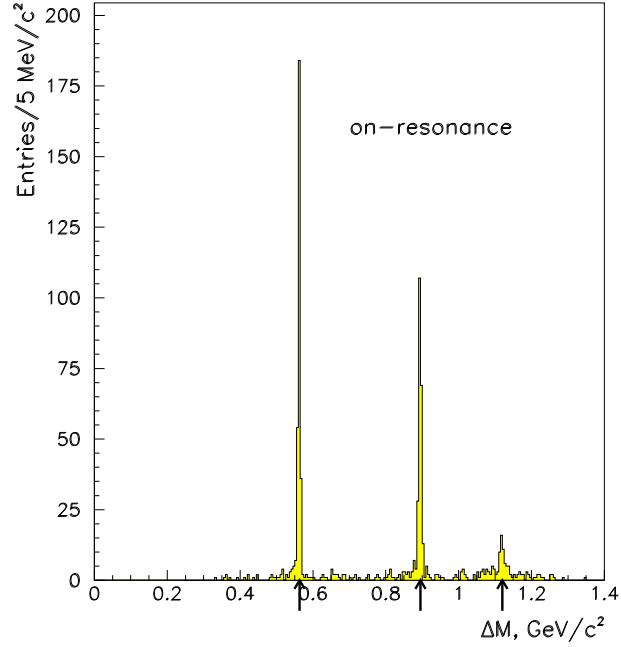}
\caption{The mass difference $\Delta M = (M_{\mu^+\mu^-\pi^+\pi^-} - 
M_{\mu^+\mu^-})$ distribution where $M_{\mu^+\mu^-}$  lies in the
$\Upsilon(1S)$ mass region.
Arrows show the positions 
of the mass differences ($M_{\Upsilon(2S)}-M_{\Upsilon(1S)}$), 
($M_{\Upsilon(3S)}-M_{\Upsilon(1S)}$) and ($M_{\Upsilon(4S)}-M_{\Upsilon(1S)}$)
from PDG~\cite{PDG}, respectively.}
\label{fg03}
\end{figure}

\newpage
\begin{minipage}[b]{8.cm}
\vspace*{.5cm} 
\includegraphics[width=8.cm]{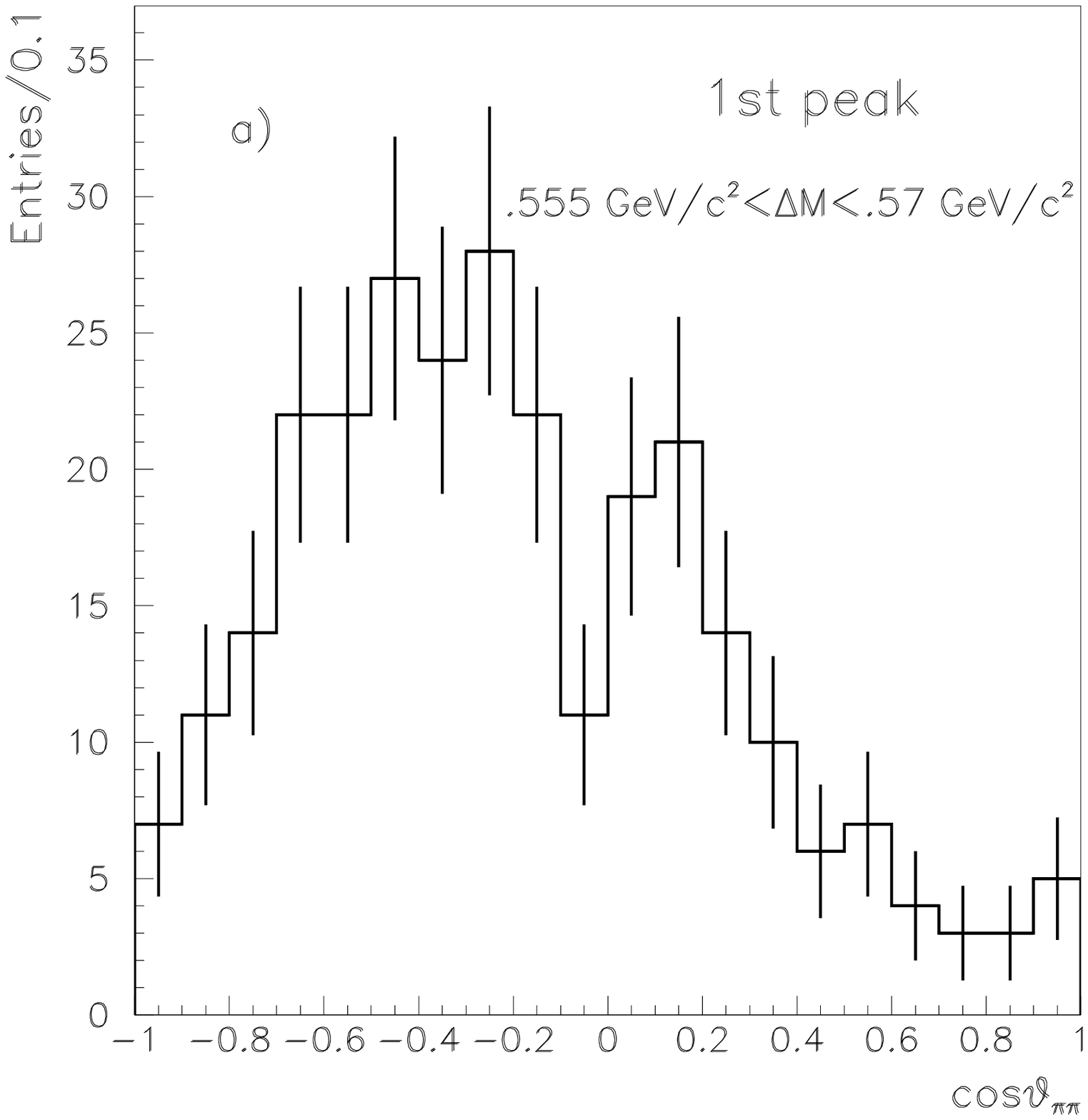}
\end{minipage}
\begin{minipage}[b]{8.cm}
\vspace*{.5cm} 
\includegraphics[width=8.cm]{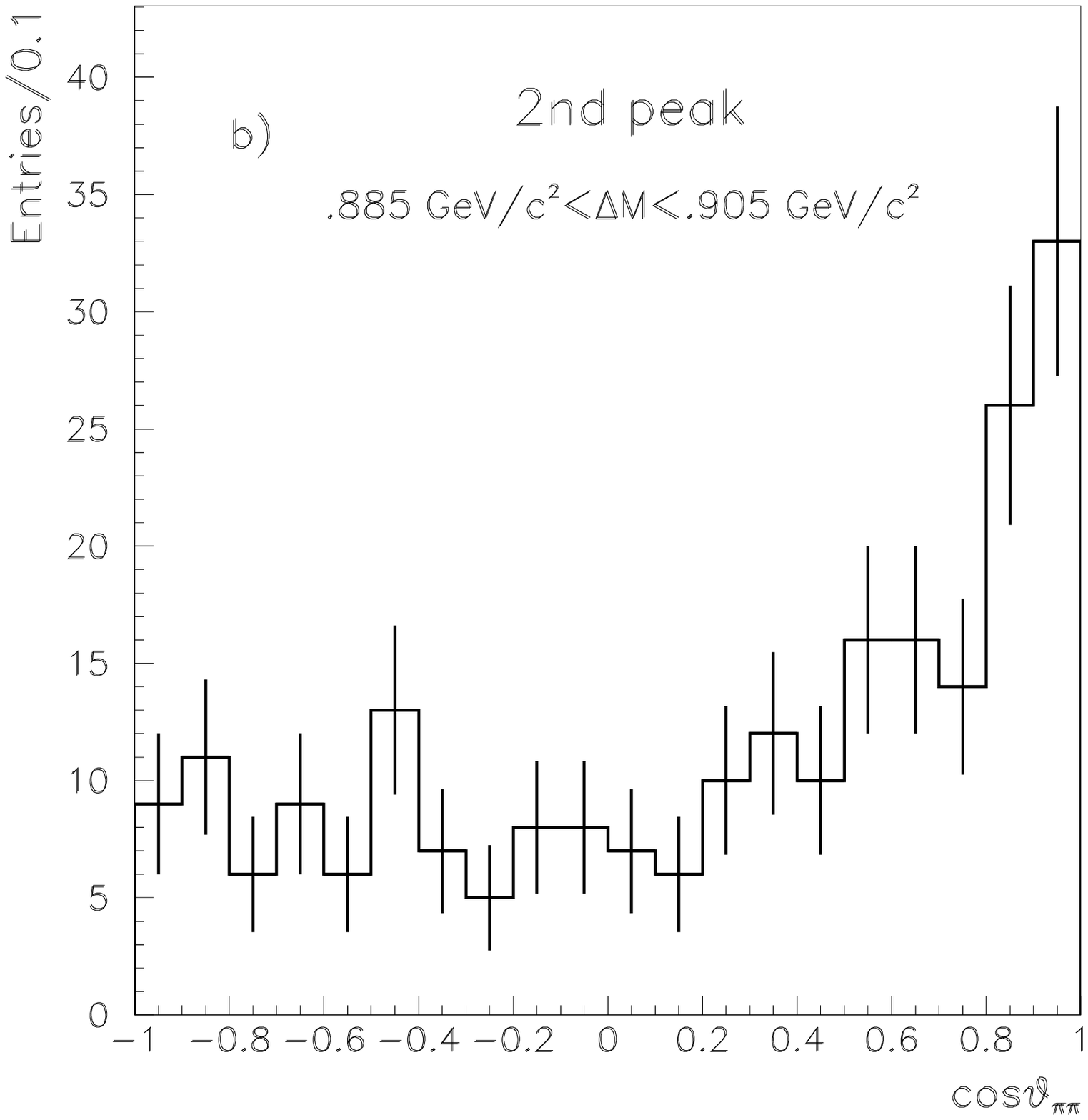}
\end{minipage}
\begin{minipage}[b]{8.4cm}
\vspace*{.5cm} 
\includegraphics[width=8.4cm]{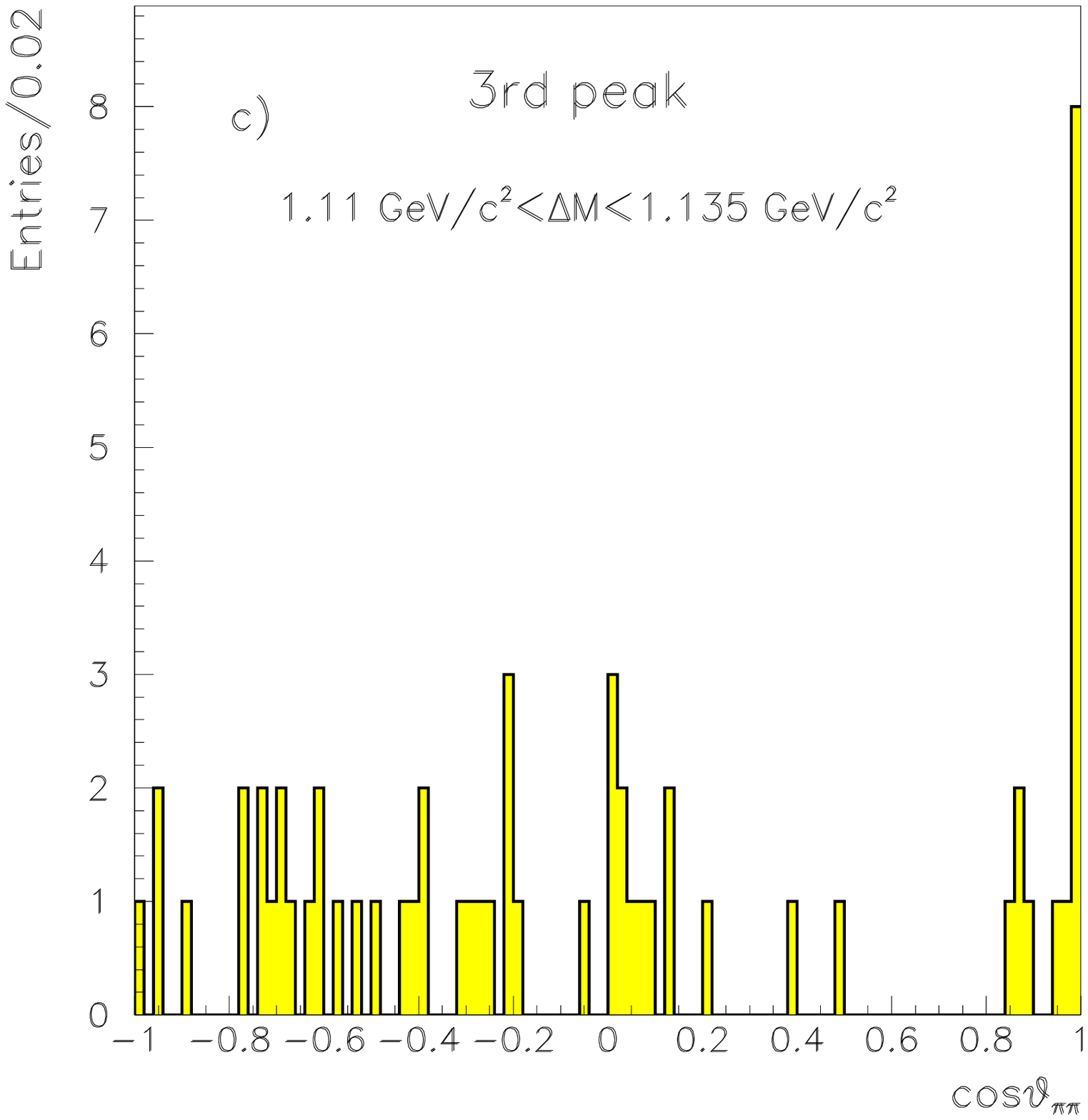}
\end{minipage}
\begin{minipage}[b]{8.4cm}
\vspace*{.5cm} 
\includegraphics[width=8.4cm]{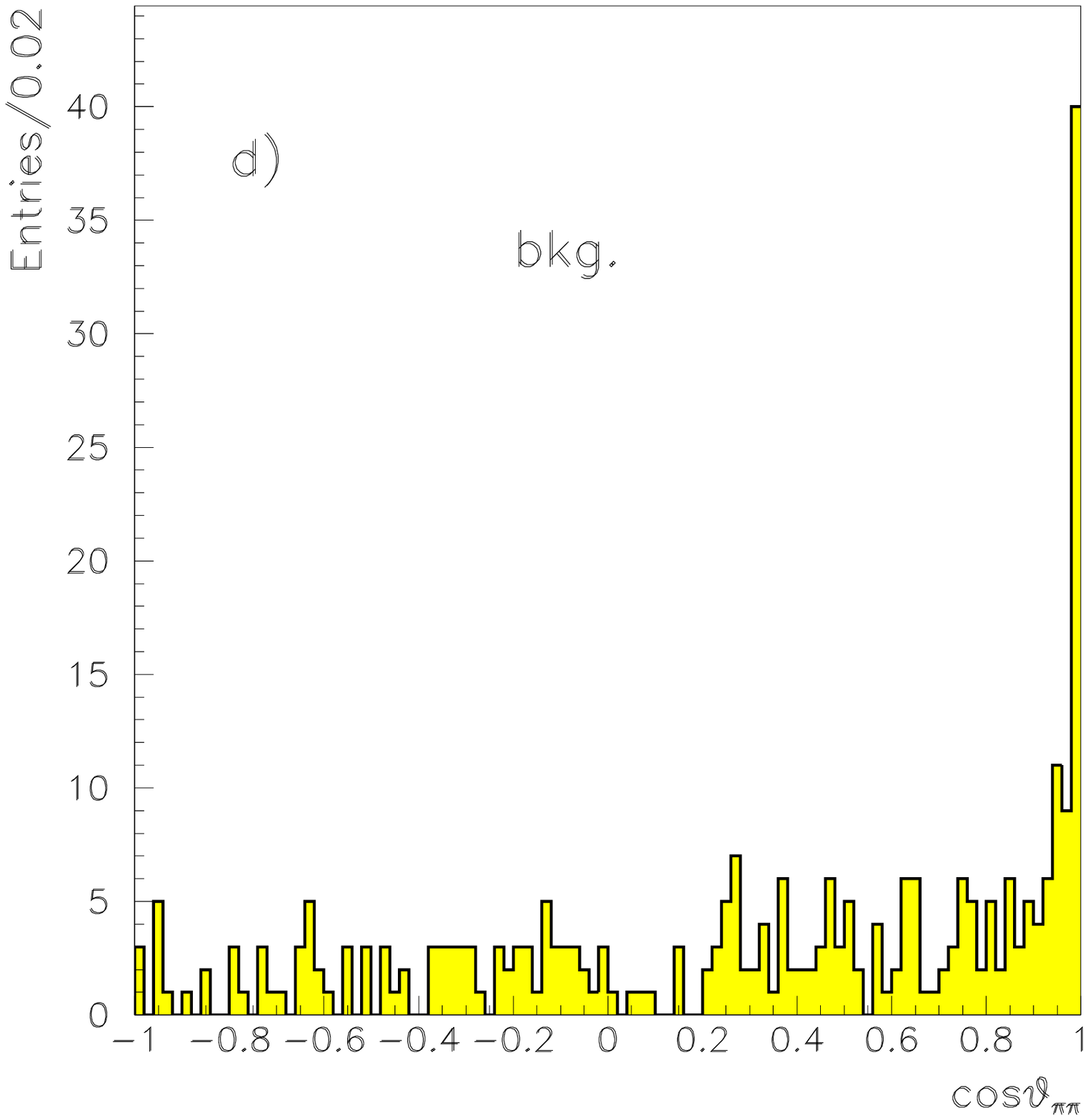}
\end{minipage}
\begin{figure}[!htb]
\vspace*{-.8cm} 
\caption{The cos$\theta_{\pi\pi}$ distributions for events from the three 
observed peaks and background.} 
\label{fg04}
\end{figure}

\newpage
\begin{minipage}[b]{7.9cm}
\vspace*{.5cm} 
\includegraphics[width=6.7cm]{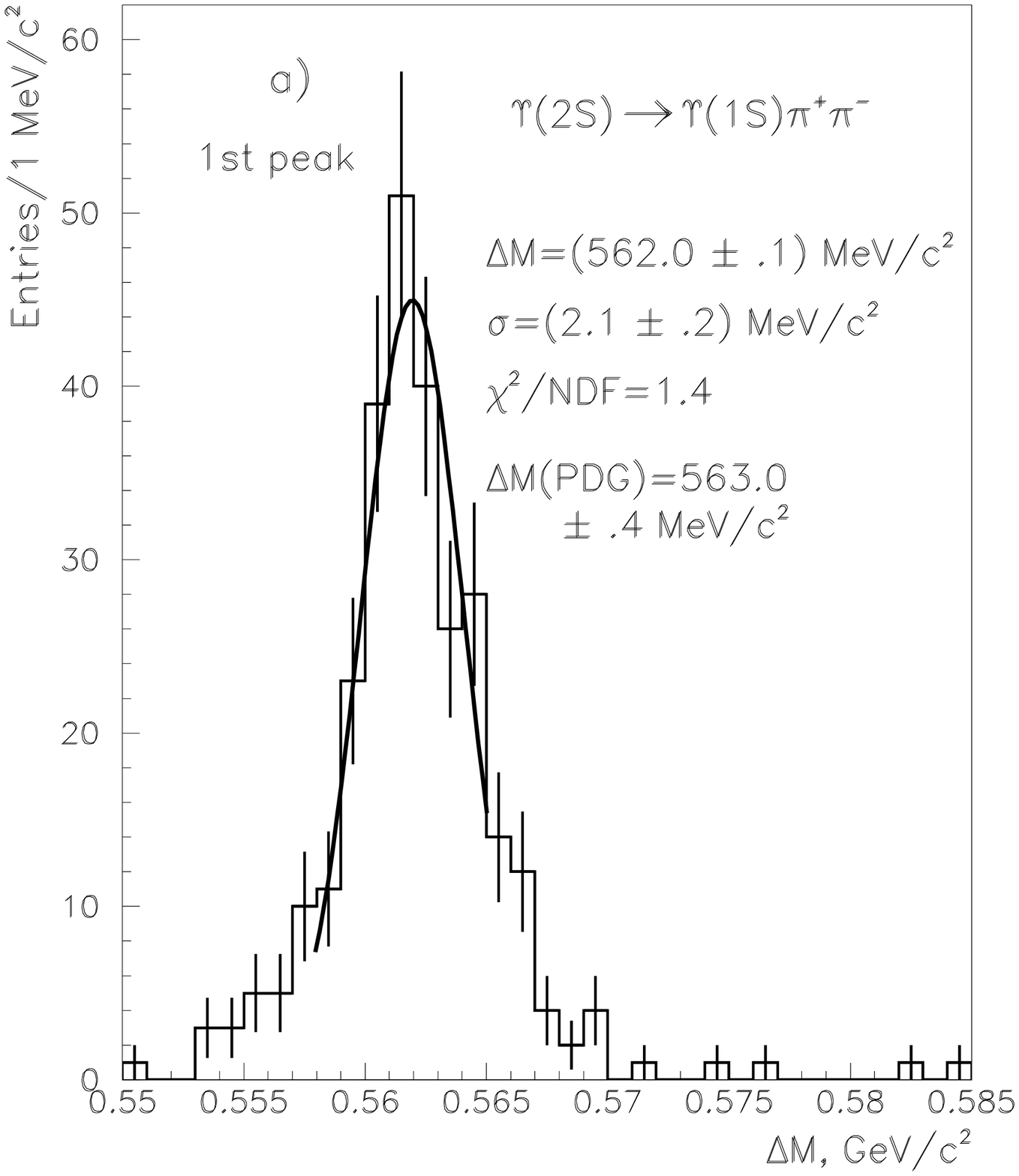}
\end{minipage}
\begin{minipage}[b]{7.9cm}
\vspace*{.5cm} 
\includegraphics[width=6.7cm]{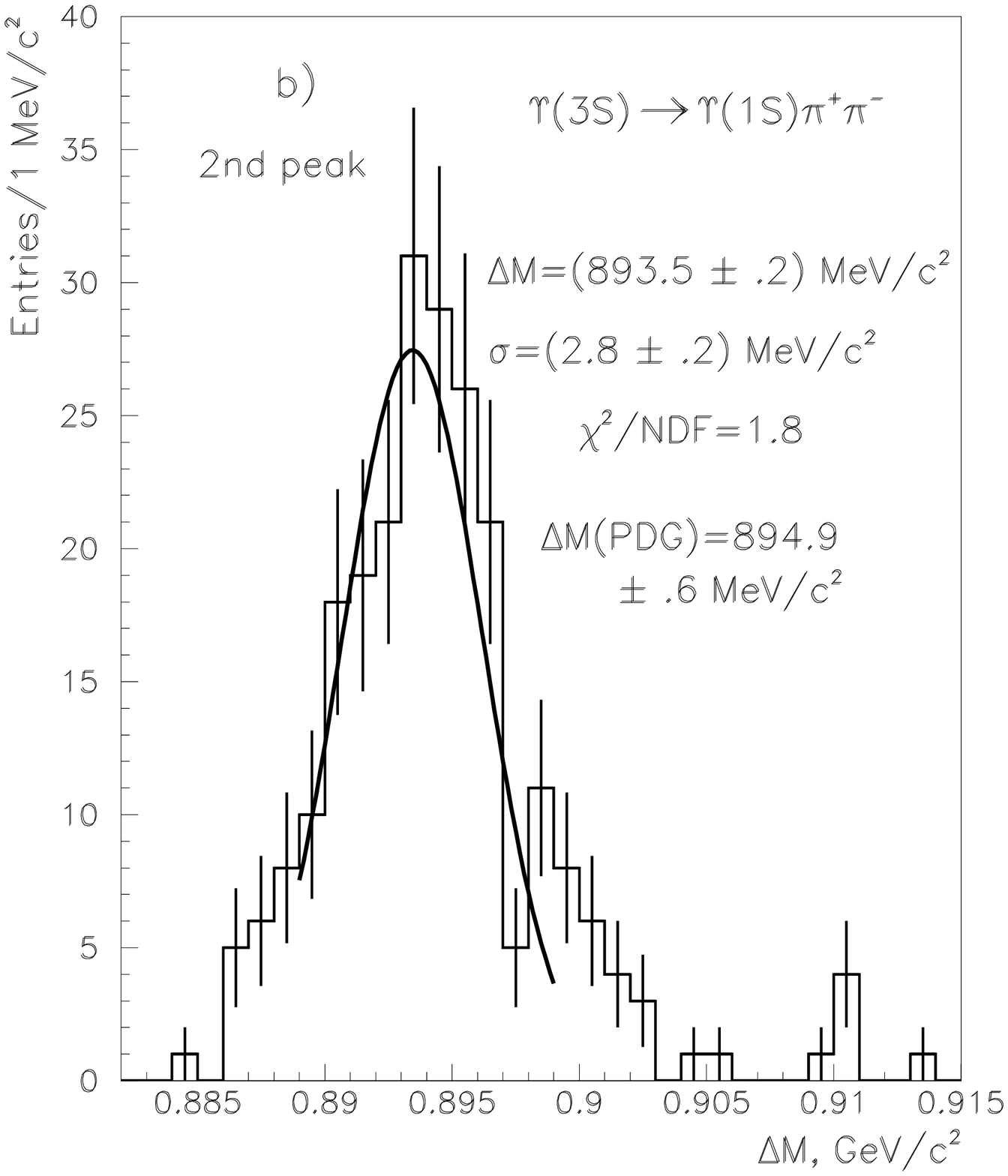}
\end{minipage}
\begin{figure}[!htb]
\caption{The mass difference $\Delta M = (M_{\mu^+\mu^-\pi^+\pi^-} - 
M_{\mu^+\mu^-})$ distribution for the first (a) and second (b) peak 
($|M_{\mu^+ \mu^-}-M_{\Upsilon(1S)}|\!<$6 MeV/$c^2$).
The central parts of the peaks are fitted by Gaussian functions.}
\label{fg05}
\end{figure}

\begin{figure}[!htb]
\includegraphics[width=0.44\textwidth]{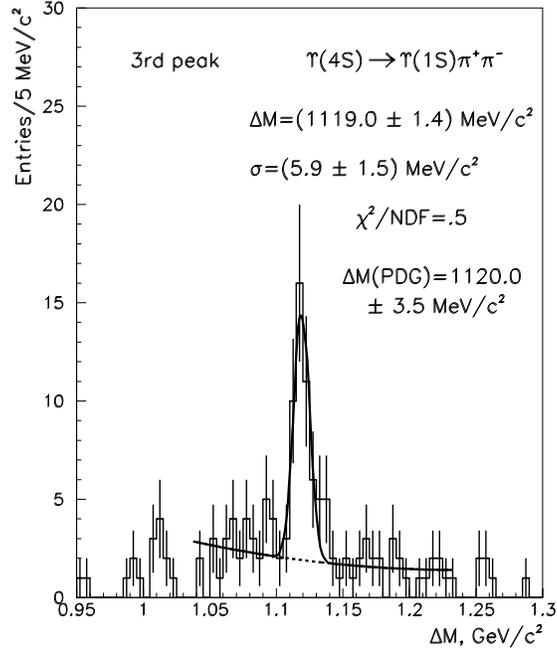}
\caption{The fit to the third peak in the mass difference 
$\Delta M = (M_{\mu^+\mu^-\pi^+\pi^-} - M_{\mu^+\mu^-})$ distribution 
($|M_{\mu^+ \mu^-}-M_{\Upsilon(1S)}|\!<$6 MeV/$c^2$) using the sum of a
Gaussian and polynomial function.}
\label{fg06}
\end{figure}

\begin{figure}[!htb]
\includegraphics[width=0.45\textwidth]{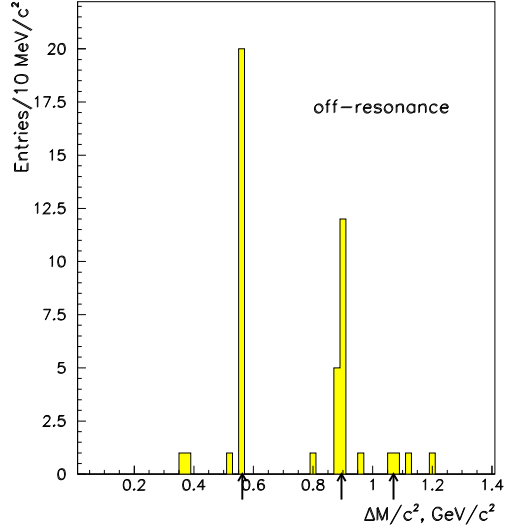}
\caption{The mass difference $\Delta M = (M_{\mu^+\mu^-\pi^+\pi^-} - 
M_{\mu^+\mu^-})$ distribution for the off-resonance 
$\mu^+\mu^-\pi^+\pi^-X$-sample
($|M_{\mu^+ \mu^-}-M_{\Upsilon(1S)}|\!<$6 MeV/$c^2$).}
\label{fg07}
\end{figure}

\vspace*{-1.2cm} 
\begin{minipage}[b]{7.6cm}
\includegraphics[width=6.6cm]{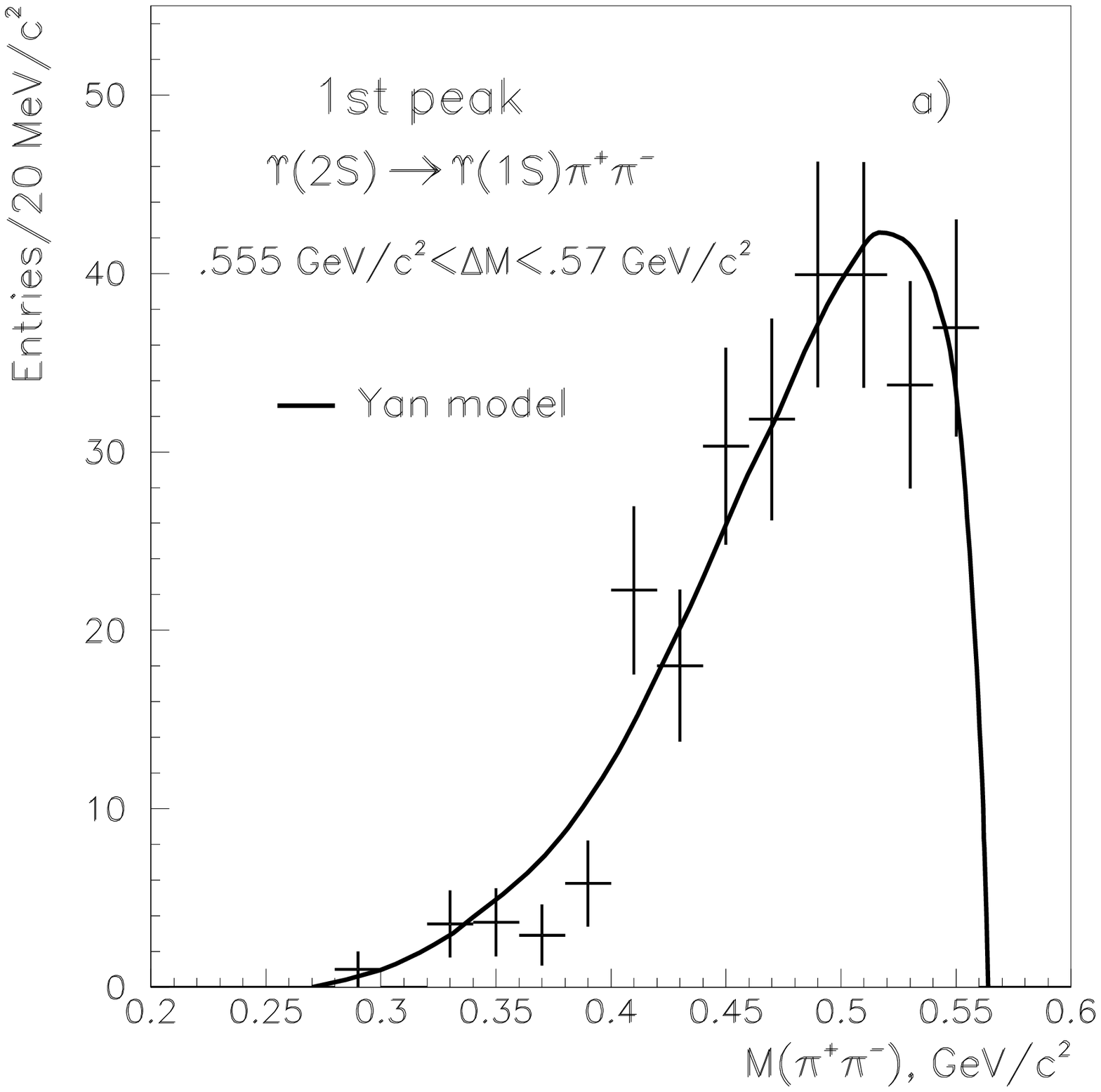}
\end{minipage}
\begin{minipage}[b]{10.5cm}
\includegraphics[width=6.6cm]{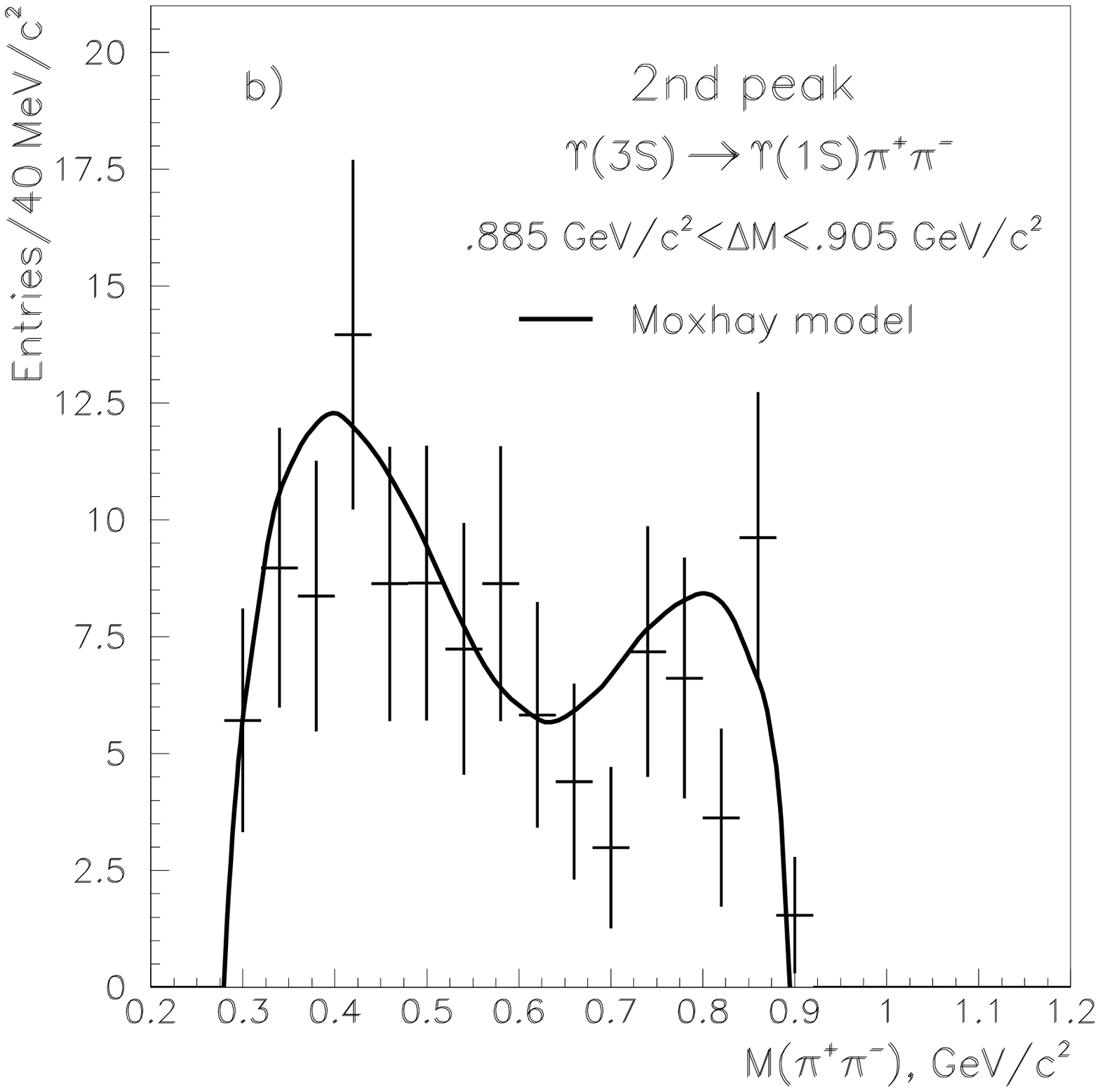}
\end{minipage}
\begin{minipage}[b]{8.cm}
\includegraphics[width=6.8cm]{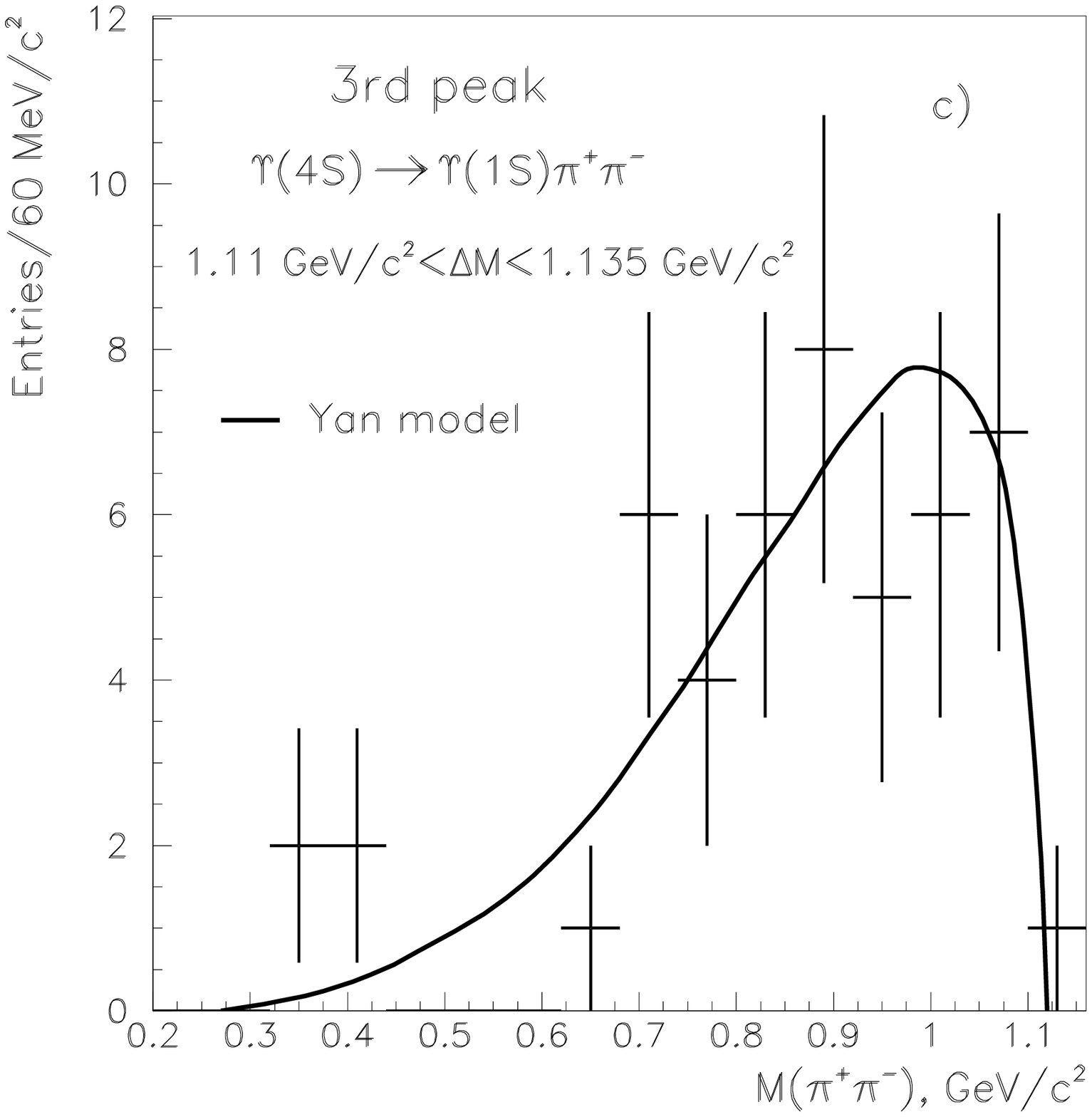}
\end{minipage}
\begin{minipage}[b]{8.cm}
\includegraphics[width=6.8cm]{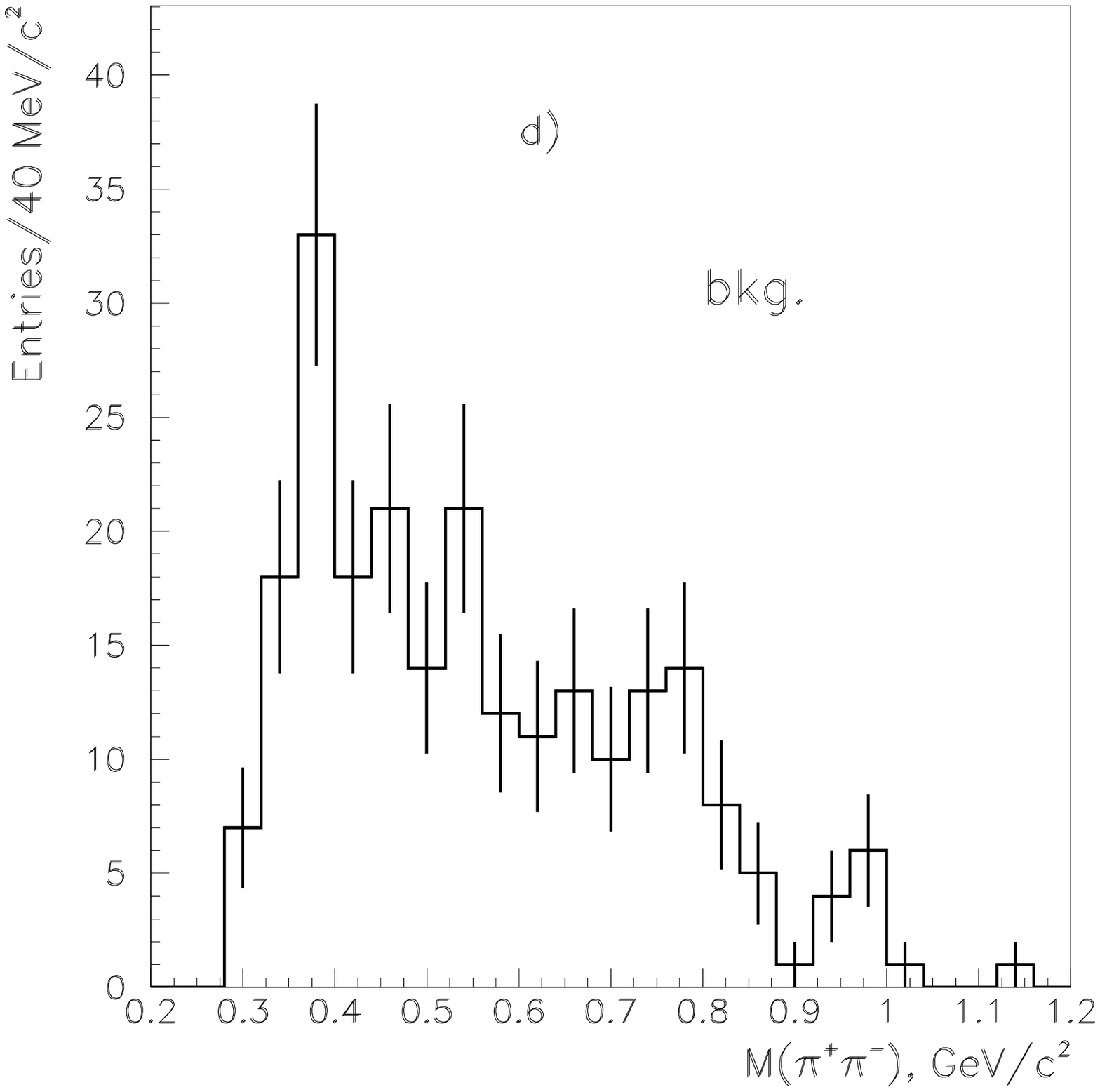}
\end{minipage}
\begin{figure}[!htb]
\vspace*{-.8cm} 
\caption{The $\pi^+\pi^-$ invariant mass distributions
         for events from the observed peaks and background.}
\label{fg08}
\end{figure}

\end{document}